\documentclass[aps,floatfix,superscriptaddress,twocolumn,amsmath,amssymb,nofootinbib,10pt,longbibliography,pra]{revtex4-2}

\usepackage{braket}
\usepackage{fontawesome}
\usepackage{graphicx}
  \graphicspath{{figures/}}
\usepackage{hyperref}
  \hypersetup{colorlinks=true,linkcolor=blue,citecolor=blue,urlcolor=blue}
\usepackage{mathtools}
\usepackage[separate-uncertainty=true]{siunitx}
\usepackage[caption=false]{subfig}
\begin{document}

\title{
Guarantees on the structure of experimental quantum networks
}

\author{Andrés Ulibarrena}
\thanks{andresulibarrena\faAt{}tuta.com}
\affiliation{Institute of Photonics and Quantum Sciences, School of Engineering and Physical Sciences, Heriot-Watt University, Edinburgh EH14 4AS, United Kingdom}

\author{Jonathan W. Webb}
\affiliation{Institute of Photonics and Quantum Sciences, School of Engineering and Physical Sciences, Heriot-Watt University, Edinburgh EH14 4AS, United Kingdom}

\author{Alexander Pickston}
\affiliation{Institute of Photonics and Quantum Sciences, School of Engineering and Physical Sciences, Heriot-Watt University, Edinburgh EH14 4AS, United Kingdom}

\author{Joseph Ho}
\affiliation{Institute of Photonics and Quantum Sciences, School of Engineering and Physical Sciences, Heriot-Watt University, Edinburgh EH14 4AS, United Kingdom}

\author{Alessandro Fedrizzi}
\affiliation{Institute of Photonics and Quantum Sciences, School of Engineering and Physical Sciences, Heriot-Watt University, Edinburgh EH14 4AS, United Kingdom}

\author{Alejandro Pozas-Kerstjens}
\thanks{physics\faAt{}alexpozas.com}
\affiliation{Group of Applied Physics, University of Geneva, 1211 Geneva 4, Switzerland}
\affiliation{Instituto de Ciencias Matem\'aticas (CSIC-UAM-UC3M-UCM), 28049 Madrid, Spain}

\begin{abstract}
Quantum networks connect and supply a large number of nodes with multi-party quantum resources for secure communication, networked quantum computing and distributed sensing.
As these networks grow in size, certification tools will be required to answer questions regarding their properties.
In this work we demonstrate a general method to guarantee that certain correlations cannot be generated in a given quantum network.
We apply quantum inflation methods to data obtained in quantum group encryption experiments, guaranteeing the impossibility of producing the observed results in networks with fewer optical elements.
Our results pave the way for scalable methods of obtaining device-independent guarantees on the network structure underlying multipartite quantum protocols.
\end{abstract}

\maketitle

As quantum information processing matures, there is an increased demand for certifying hardware devices or the underlying quantum resource with minimal assumptions.
It is now possible to guarantee quantum phenomena from just statistics corresponding to few, uncharacterized measurements, using the device-independent framework \cite{acin2007di,brunner2014bell}.
Despite the amount of information from the system being minimal, the device-independent formalism allows, in certain situations, not just to guarantee that the device under scrutiny is quantum but to certify the quantum state and measurements that are being performed on it \cite{supic2020selftesting}, and properties such as nonlocality \cite{brunner2014bell}, entanglement \cite{baccari2017di,baccari2020genuinely}, randomness \cite{borkala2022randomness}, the dimension of the underlying quantum state \cite{brunner2008dimension}, quantum measurements \cite{rabelo2011di,smania2019di}, or superpositions of causal orders \cite{vanderlugt2022}.

The focus of device-independent quantum certification has been moving towards complex networks that feature several independent quantum systems being distributed to multiple parties \cite{tavakoli2021bell}.
In analogy with the device-independent certification methods in bipartite scenarios---which follow the spirit of Bell's theorem \cite{bell1964einstein}---many Bell-like inequalities have been developed that are satisfied by all correlations that can be generated in specific networks when the sources distribute classical systems \cite{branciard2010characterizing,branciard2012bilocal,fraser2018triangle,luo2018networks,rosset2016networks,tavakoli2014nonlocal,mukherjee2015,tavakoli2016connected,pozas2022full,luo2023hierarchical}, and some of them have been found not to be satisfied in experiments \cite{saunders2017experimental,carvacho2017experimental,sun2019experimental,poderini2020experimental,baumer2021}.
Moreover, the device-independent analysis of the limits of quantum mechanics in networks has led to the demonstration of the necessity of complex numbers in order to account for all quantum correlations observed in nature \cite{li2021real,chen2021real}.

In this work, we take an orthogonal approach to certification in quantum networks.
Typically, one assumes that the network structure underlying the setup under study is fixed, and non-compliance with the corresponding constraints signal the use of supra-classical \cite{branciard2010characterizing,fritz2012beyond,renou2019genuine} or supraquantum \cite{pozas2023postquantum} resources.
In contrast, we focus on providing guarantees on the network structure underlying some observations.
As quantum networks become commonplace and more complex, it is vital to develop efficient and scalable tools that allow users to guarantee their integrity, including the network structure itself (which could be modified by colluding parties or by eavesdroppers), in order to correctly perform quantum protocols on them, safe from eavesdrops.
In fact, it is already known that being able to guarantee a particular network structure allows certain, otherwise impossible \cite{Mayers97}, cryptographic protocols \cite{Luo2020b}.

In order to provide such certifications on the structure of quantum networks, we assume that quantum mechanics accurately describes natural phenomena and thus there are no supraquantum resources in nature.
In this case, observing a behavior beyond the limits of what is allowed by quantum mechanics in a given network is a demonstration that the actual network that is implemented is a different one.
In this sense, the goal of our work aligns with that of Refs.~\cite{coiteux2021prl,coiteux2021pra,cao2022exp,mao2022exp}, having the additional hypothesis that quantum mechanics accurately describes nature.
The recent work \cite{weinbrenner2023} addresses this problem under the additional assumptions that the sources distribute states (close to) Greenberger-Horne-Zeilinger states and each party receives just one qubit.
In contrast, the approach we describe in this work only assumes the validity of quantum mechanics and the independence of the sources of quantum systems in the network, thus taking a full device-independent perspective.

\begin{figure*}
    \includegraphics[width=\linewidth]{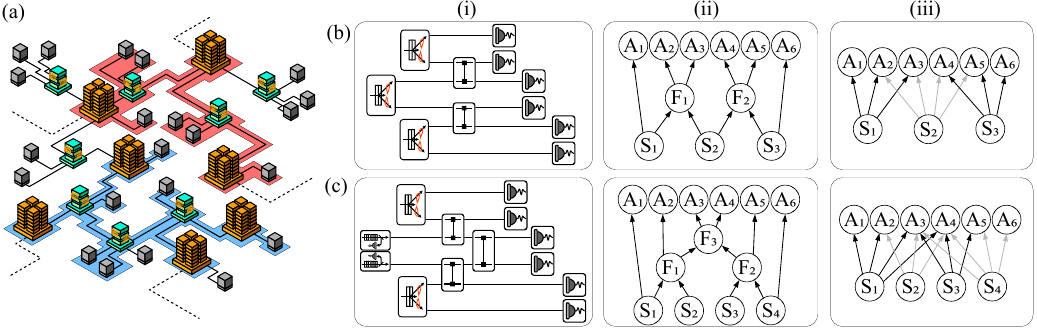}
    \caption{Testing the network structure in uncharacterized realizations of quantum networks.
    (a) A prototype quantum network: the golden, tall buildings represent factories of quantum states, which are distributed either to end-users (gray, small detectors) or ``data centers'' (green, medium buildings) that are allowed to perform arbitrary joint transformations on all the systems received).
    Our goal is to test the network structure just from observational data in the gray nodes.
    We illustrate the method in two networks: The process for the red shaded network is elaborated in row (b) and detailed in the main text, while the network in the blue region is discussed in row (c) and in Appendix~\ref{app:trident}.
    Column (i) depicts specific realizations implemented in the corresponding networks, using circuit notation (i.e., with all the sources, gates and measurements).
    Column (ii) depicts the causal diagrams corresponding to the networks, where the sources $S_i$ send systems to interact with each other in the nodes $F_i$, and subsequently are distributed to the parties $A_i$ which measure them.
    Column (iii) depicts network structures (with states and measurements, but no operations) that will be used to attempt to reproduce the correlations generated in the structures in column (ii).
    Our procedure consists in finding the most general network (column (iii)) with as many sources and parties as the realization (columns (i)-(ii)), characterizing the observations that can be produced in it, and demonstrating that the measurement statistics produced in the experiment do not belong to this set.
    }
    \label{fig:NQKDNetwork}
\end{figure*}

Our approach consists of, given some observations coming from an uncharacterized realization (e.g., columns (i-ii) in Fig.~\ref{fig:NQKDNetwork}) proposing candidate network structures (e.g., column (iii) in Fig.~\ref{fig:NQKDNetwork}), and test whether the candidate structure can generate the observations.
Our goal will be to prove that the observations are incompatible with the candidate network, this is, that there exist no quantum states (of any dimension) and measurements that, once distributed according to the network structure, reproduce the observations.

We illustrate this approach by analyzing data from the realization of six-photon experiments~\cite{pickston_nqkdtrident_2022,webb2023} used in quantum conference key agreement (QCKA) protocols~\cite{epping_large-scale_2016,epping_multi-partite_2017,hahn_quantum_2019,wallnofer2019,murta2020quantum, proietti_experimental_2021}.
We do so for two reasons:
First, quantum cryptography is an important application where, if security proofs assume a specific network structure, guaranteeing this structure experimentally becomes crucial to discard potential eavesdrops \cite{lee2018towards}.
Second, device-independent certification in QCKA protocols is achieved by contrasting the observed statistics against multipartite global local hidden variable models or by certifying the presence of genuine multipartite entanglement \cite{ribeiro2018,philip2023}.
However, if the protocol is implemented in a given network, contrasting against the corresponding network models has advantages in terms of the requirements for certification (see, e.g., \cite{mukherjee2015,pozas2019activation} for advantages in terms of the detection efficiency) in addition to those discussed above.
Using readily available tools \cite{wolfe2021qinflation,inflapyon}, we will produce quantum Bell-like inequalities for particular hypotheses on the network structure underlying the experimental implementations and observe their violation by the empirical statistics.

\section*{Necessary constraints on quantum network correlations}
Characterizing the correlations that are generated in network scenarios is a notably hard problem.
For some networks, there exist simple necessary conditions for correlations to be compatible: when two parties share no causal history, their joint distribution factorizes.
This is the case, for instance, of parties $A_1$ and $A_6$ in Figs.~\ref{fig:NQKDNetwork}(b-ii) and~(b-iii), or between the extremal parties in the entanglement-swapping network \cite{branciard2010characterizing}.
However, there exist networks where no such factorizations appear.
The simplest example is the triangle network, obtained from the entanglement-swapping network by adding a new source connecting the extremal parties.
This network, being the simplest one where explicit factorizations fail to characterize it, has been subject to intense study \cite{fraser2018triangle,gisin2020constraints,abiuso2021,boreiri2022,pozas2023,pozas2023postquantum}.

In these cases without explicit factorizations, one can derive necessary conditions for correlations compatible with a network by means of inflation~\cite{wolfe2019inflation,wolfe2021qinflation}.
Briefly, inflation allows to derive compatibility constraints by imagining that multiple copies of the sources distributing physical systems and of the measurement devices held by the parties are available, and analyzing the correlations that are obtained when connecting these copies.
The network structure is reflected in symmetries in the correlations on the new, inflated networks, which are much simpler to enforce and analyze (see Appendix~\ref{app:inflation}).
Indeed, many of such constraints are now present in the literature, mostly in the form of Bell-like inequalities \cite{fraser2018triangle,gisin2020constraints,pozas2023}.
The inequalities provided by inflation methods are polynomial, i.e. of the form
\begin{equation}
    \sum_n \sum_{\substack{\vec{a}_1\dots\vec{a}_n\\\vec{x}_1\dots\vec{x}_n}} c_{\vec{a}_1,\dots,\vec{a}_n,\vec{x}_1,\dots,\vec{x}_n} p(\vec{a}_1|\vec{x}_1)\cdots p(\vec{a}_n|\vec{x}_n)\geq 0,
    \label{eq:witness}
\end{equation}
where $c_{\vec{a}_1,\dots,\vec{a}_n,\vec{x}_1,\dots,\vec{x}_n}$ are real coefficients, each $\vec{a}_i$ is a vector of outputs obtained by all parties, and $\vec{x}_i$ is the vector of corresponding inputs.
These inequalities are derived naturally from the separating hyperplanes that appear when solving the linear or semidefinite programs associated to, respectively, classical \cite{wolfe2019inflation} and quantum \cite{wolfe2021qinflation} inflation problems in any network.
Finding for some $p(\vec{a}|\vec{x})$ that the left-hand side evaluates to a negative quantity is a guarantee that $p(\vec{a}|\vec{x})$ does not admit the quantum inflation used to produce the inequality.
Importantly, admitting an inflation of a network is a relaxation of admitting a realization in said network.
Therefore, detecting that some correlations do not admit an inflation of a network is a proof that they cannot be produced in the original network.

In order to illustrate the full method that we propose, let us consider the experimental realization that is used to produce six-partite Greenberger-Horne-Zeilinger (GHZ) states in Ref.~\cite{webb2023}, illustrated in Fig.~\ref{fig:NQKDNetwork}(b).
There, six-photon multipartite entangled states are created and distributed to six parties, $A_1,\dots,A_6$.
Three entangled photon-pair sources generate Bell pairs, and subsequently two fusion gates are used to obtain the final six-photon GHZ state.
After the measurements, the outcome statistics follow the corresponding Born's rule, namely
\begin{equation}
    \begin{aligned}
        p(a_1,&\dots,a_6)\\
        &=\text{Tr}\left[U_{23}\otimes U_{45}\left(\phi^+_{12}\otimes\phi^+_{34}\otimes\phi^+_{56}\right)U^\dagger_{23}\otimes U^\dagger_{45}\right.\\
        &\left.\quad\qquad\left(\Pi_{a_1}\otimes\cdots\otimes\Pi_{a_6}\right)\right],
    \end{aligned}
    \label{eq:expstatistics}
\end{equation}
where $\phi^+=\frac12(\ket{00}+\ket{11})(\bra{00}+\bra{11})$ is the maximally entangled state, $U_{ij}$ is the unitary implementing the fusion gate between photons $i$ and $j$, and $\Pi_{a_i}$ are the projectors describing the measurement operator of party $A_i$.
One can consider distributions with inputs, $p(a_1,\dots,a_6|x_1,\dots,x_6)$ by using different projectors $\Pi_{a_i}^{x_i}$ for each measurement.
For more details on the six-photon state generation, we refer to Appendix~\ref{app:experimental} and the original reference \cite{webb2023}.

The first step in the procedure is finding a network (i.e., a bipartite graph representation that only contains sources and parties \cite{tavakoli2021bell}) that closely resembles the structure of the experiment.
This network will have as many sources and outcomes as the experimental realization.
Each of the sources will distribute systems to all the parties which, in the experiment, have a causal connection to it.
For the setup of Fig.~\ref{fig:NQKDNetwork}(b) this means that the leftmost (respectively rightmost) source will distribute systems to the three leftmost (respectively rightmost) parties, and the central source will distribute systems to the four central parties, leading to the network in Fig.~\ref{fig:NQKDNetwork}(b-iii).
Distributions that are generated in this network take the form given by the corresponding Born's rule, i.e.,
\begin{equation}
    \begin{aligned}
        p(a_1,&\dots,a_6)\\
        &=\text{Tr}\left[\left(\rho_{S_1}\otimes\rho_{S_2}\otimes\rho_{S_3}\right)\cdot\left(\Pi_{a_1}\otimes\cdots\otimes\Pi_{a_6}\right)\right],
    \end{aligned}
    \label{eq:networkstatistics}
\end{equation}
where each $\rho_{S_i}$ represents an arbitrary state distributed by source $S_i$.

In order to discern whether the quantum correlations generated in the structure in Figs.~\ref{fig:NQKDNetwork}(b-i), (b-ii) via Eq.~\eqref{eq:expstatistics} can be reproduced in the network of Fig.~\ref{fig:NQKDNetwork}(b-iii) via Eq.~\eqref{eq:networkstatistics} we will use quantum inflation \cite{wolfe2021qinflation} as described in Appendix~\ref{app:inflation} (namely, by relaxing the problem to a hierarchy of semidefinite programs that test the existence of distributions on extended scenarios with appropriate symmetries and constraints over their marginals).
This implies, in particular, that we impose no restriction on the dimension of the systems distributed by the sources nor in the measurements that the parties perform on all the shares of their respective systems.
We thus allow to create strong correlations between the systems in the network in Fig.~\ref{fig:NQKDNetwork}(b-iii).
Yet, we will show that these are not strong enough to reproduce the multi-photon correlations observed in Fig.~\ref{fig:NQKDNetwork}(b-i).

In the remainder of the manuscript we focus on analyzing conditions that correlations that can be generated in the network of Fig.~\ref{fig:NQKDNetwork}(b-iii) and how the experimental data produced in Fig.~\ref{fig:NQKDNetwork}(b-i) (found in Ref.~\cite{webb2023}) does not meet them, showcasing the importance of the fusion gates in the realization.
We must stress that our approach is fully general, not restricted to the setup in Fig.~\ref{fig:NQKDNetwork}(b-i).
In order to illustrate the generality of the approach, in Appendix~\ref{app:trident} we perform an analogous analysis for the experiment carried out in Ref.~\cite{pickston_nqkdtrident_2022}, depicted in Fig.~\ref{fig:NQKDNetwork}(c).

\section*{Witnesses of network incompatibility}
When using quantum inflation, witnesses of incompatibility can be obtained in a direct manner, exploiting the fact that the compatibility of a distribution with a quantum inflation can be formulated as a semidefinite program \cite{wolfe2021qinflation,tavakoli2023semidefinite}.
These are optimization problems that, upon finding an incompatible distribution (recall, one for which no quantum states and measurement operators exist that reproduce it in the candidate network), provide a witness in the form of Eq.~\eqref{eq:witness} that is positive for all compatible distributions and evaluates negatively at least for the incompatible one.
Importantly, its evaluation to a negative quantity by any distribution is a guarantee that such distribution does not admit a realization in the candidate network.

We will obtain these witnesses for several distributions of the form of Eq.~\eqref{eq:expstatistics}.
Then, as a second step, we will evaluate the witnesses on the empirical data obtained in Ref.~\cite{webb2023}.
In order to do so, we convert the raw counts from the detectors into a probability distribution of six-photon events.
The experimental setup employs measurement stages with two outputs, denoted by the transmission of a horizontally polarised photon through the polarising beam splitter or the reflection of a vertically polarised one.
By normalising the number of six-photon counts (one per party) obtained in each of the possible events by the total number of six-photon counts, we obtain the empirical distributions that we will test.

\subsection*{Binary-input distribution}
The data in Ref.~\cite{webb2023} contains counts for all the parties measuring in the $X$ and $Z$ bases.
Therefore, it is possible to consider the two-input distribution $p(a_1,\dots,a_6|x_1,\dots,x_6)$, where $x_i=0$ corresponds to the measurement on the $X$ basis and $x_i=1$ corresponds to the measurement on the $Z$ basis.
We find the resulting theoretical distribution \eqref{eq:expstatistics}, using ideal states and measurements, not to admit a realization in the network of Fig.~\ref{fig:NQKDNetwork}(b-iii) (i.e., an expression of the form of Eq.~\eqref{eq:networkstatistics}) by using its corresponding second-order inflation (depicted in Fig.~\ref{fig:inflation} in Appendix~\ref{app:inflation}) and, already, at the first level of the associated Navascu\'es-Pironio-Ac\'in (NPA) hierarchy \cite{npa1,npa2}.
The hierarchy is defined via sets of operators $\mathcal{O}_n$ that index the rows and columns of the matrix $\Gamma^n_{i,j}=\text{Tr}[\rho\cdot O_i^\dagger O_j]$.
If a distribution admits a quantum realization, $\Gamma^n$ is positive semidefinite for any generating set $\mathcal{O}_n$, and if it does not there exists at least one $\mathcal{O}_n$ for which $\Gamma^n$ is negative definite.
The first level of the hierarchy, which we use for analyzing the two-input distribution, is defined by the set of operators $\mathcal{O}_1:=\{\openone\}\cup\{A^{i,j}_p\}$, leading to a matrix of size $41\times 41$ in our case of interest, whose positivity can be determined in $<1$ seconds.
In Appendix~\ref{app:twoinput} we elaborate on the details of the implementation.

The guarantee of incompatibility is given by the witness in Eq.~\eqref{eq:twoinput_1} in Appendix~\ref{app:twoinput} (see also the computational appendix \cite{compapp} for the code executed to obtain it), extracted from the problem.
This witness does not only identify Eq.~\eqref{eq:expstatistics} as incompatible, but when adding white noise to the maximally entangled states, $\phi^+_{ij}\mapsto v\phi^+_{ij}+(1-v)\openone/4$, it identifies that the distribution is incompatible at least for $v\gtrsim0.6180$.
By increasing the accuracy of the approximations of Eq.~\eqref{eq:networkstatistics} provided by inflation (see Appendix~\ref{app:twoinput}) it is possible to guarantee incompatibility for, at least, all $v\gtrsim0.3887$.

\begin{figure}
    \centering
    \includegraphics[width=0.98\columnwidth]{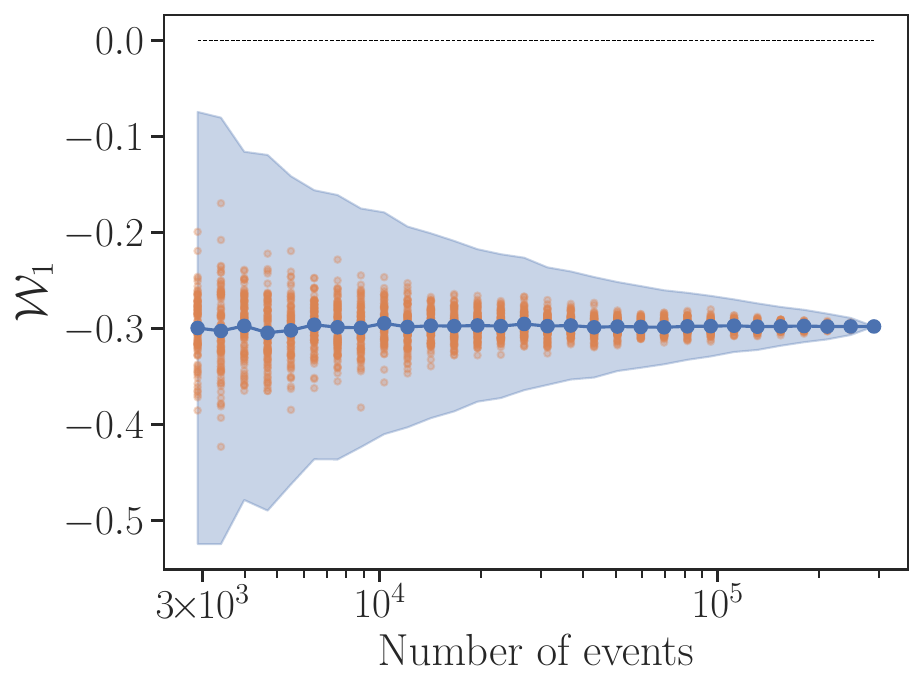}
    \caption{Evaluation of the structure witness in Eq.~\eqref{eq:twoinput_1} in the experimental data of Ref.~\cite{webb2023}.
        The maximum value achievable by quantum distributions generated in the network of Fig.~\ref{fig:NQKDNetwork}(b-iii) is upper bounded by 0, so evaluating to a negative number by a given distribution is a witness that such distribution cannot be generated in Fig.~\ref{fig:NQKDNetwork}(b-iii).
        In the horizontal axis we denote the amount of all datapoints, chosen at random, used for computing the witness.
        Error bars correspond to five standard deviations over 100 repetitions.
        The individual results are depicted by the orange points.
        The magnitude of the witness gives a notion of the distance to the set of compatible distributions, but in general it lacks of a concrete physical meaning.}
    \label{fig:twoinput}
\end{figure}

The evaluation on the experimental data is shown in Fig.~\ref{fig:twoinput}.
Notably, even the smallest amount of data considered, namely $\sim2\,900$ six-photon events, representing $1\%$ of the total amount, allows for a robust violation (namely $\mathcal{W}_1=-0.3420\pm0.0466$) well beyond the five-deviation limit.
This represents an acquisition time of only \SI{60}{s} per measurement basis, or equivalently, a total use of the experiment of $\sim1$ hour.
In combination with the key generation results of Ref.~\cite{webb2023}, this result demonstrates that it is feasible to dedicate a small amount of the data for certifying the network structure and use the rest to establish key at significantly higher rates than via concatenations of bipartite protocols.

We also analyze the data of Ref.~\cite{pickston_nqkdtrident_2022}, corresponding to the network in Fig.~\ref{fig:NQKDNetwork}(c), in Appendix~\ref{app:trident}.
In this case, the available data allows for obtaining three binary-input distributions, corresponding to the parties performing their measurement along the bases $\{X-Y, X-Z, Z-Y\}$.
We are able to obtain inequalities that witness incompatibility for all distributions for $v\gtrsim2^{-1/4}$.
However, the differences in time spent accumulating six-photon coincidences per measurement basis ($\sim$5 minutes for Ref.~\cite{pickston_nqkdtrident_2022}, totalling $\sim1\,000$ six-photon events per basis, versus $\sim$3.5 hours for Fig.~\ref{fig:NQKDNetwork}(b-i), totalling $\sim4\,500$ six-photon events per basis), reflect themselves in the fact that the empirical distributions are not witnessed incompatible in the former case.

\subsection*{Tests for no-input distributions}
\begin{figure*}
    \centering
    \subfloat[\label{fig:NoInputGHZTheo}]{
        \includegraphics[width=0.48\textwidth]{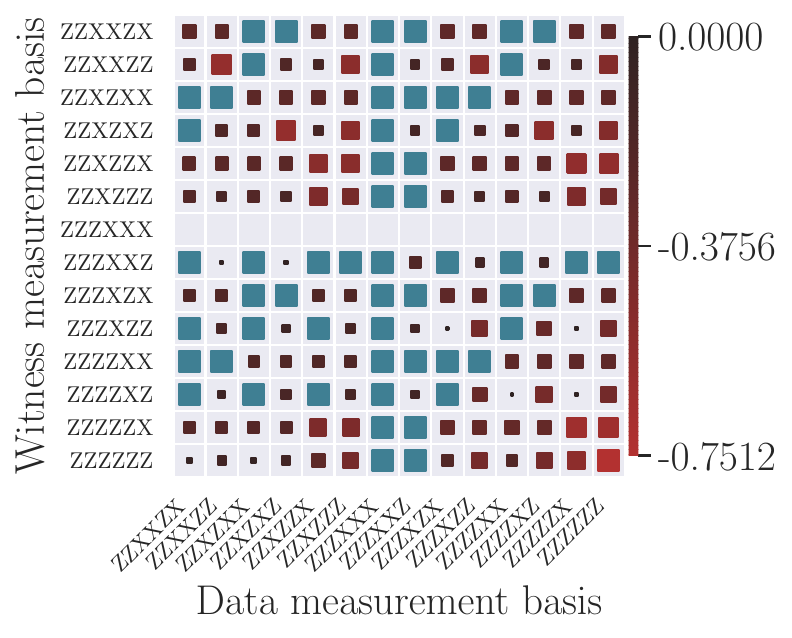}
    }
    \hfill
    \subfloat[\label{fig:NoInputGHZExp}]{
        \includegraphics[width=0.48\textwidth]{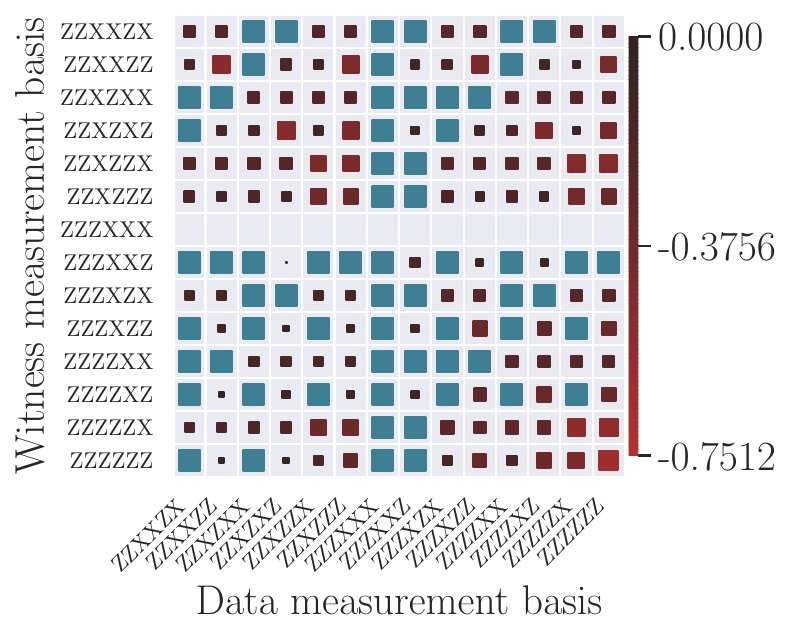}
    }
    \caption{\protect\subref{fig:NoInputGHZTheo} Theoretical predictions and \protect\subref{fig:NoInputGHZExp} experimental results for no-input witnesses of incompatibility with the network of Fig.~\ref{fig:NQKDNetwork}(b-iii).
        The indexing of the rows and columns denotes the measurement operators that are used to generate a no-input probability distribution according to Eq.~\eqref{eq:expstatistics}.
        Using the procedure described in the text, the distributions denoted by the rows are found to be incompatible with realizations in the network of Fig.~\ref{fig:NQKDNetwork}(b-iii), each one producing a witness of incompatibility.
        Then, each witness is evaluated on all distributions denoted by the columns, producing the figures where each cell represents the evaluation of the witness obtained from the distribution in the row in the distribution in the column.
        The blue cells denote distributions that are not detected by a particular witness, i.e. those that evaluate to a positive value.
        The empty rows denote ideal distributions that are not detected to be incompatible with the inflation used.
        The size and the color of the red squares denote the strength of the detection for distributions witnessed to be incompatible.
        The complete figures with the 64 possible input combinations can be found in Figs.~\ref{fig:CertificateMatrixTheoryGHZ}, \ref{fig:CertificateMatrixExpGHZ} on Appendix~\ref{app:noinput}, and a selection of the value of the inequalities as a function of the amount of data used can be found in Fig.~\ref{fig:singleinput_all}.}
    \label{fig:NoInputGHZ}
\end{figure*}

Certifying quantum properties, such as non-locality or entanglement, in a device-independent manner necessitates of the parties performing different measurements on the shares of the states received.
In contrast, it is known that constraints on the network structure are encoded even on distributions without inputs \cite{gisin2020constraints}, i.e., when the parties do not have a choice of measurement but they always measure the same operator in the received shares.
One can therefore use no-input distributions to attempt at extracting guarantees of the network structure.
This reduces the amount of data needed for the certification: while in the binary-input case one needs all the $2^{12}$ probabilities $p(a_1,\dots,a_6|x_1,\dots x_6)$ for $a_1,\dots,a_6,x_1,\dots x_6\in\{0,1\}$, using no-input distributions needs only of $n\cdot 2^6$ probabilities $\{p_k(a_1,\dots,a_6)\}_{k=1}^n$, where $n$ is the total number of distributions tested.
Therefore, in principle, one could use the techniques described earlier to give guarantees in the network structure even with fewer data.
Unfortunately, this gain does not come for free.
Any no-input distribution can always be simulated by a single source of shared randomness, and therefore any violation of a single inequality can always be attributed to an adversary classically correlating the parties' outcomes.
However, in the same way that the classical distribution $p(a,b,c)=\frac{1}{2}\text{ if }a=b=c$ can simulate the correlations of measurements on the $Z$ basis performed on the state $(\ket{000}+\ket{111})/\sqrt{2}$ but not those of $X$ measurements, having a distribution being detected by several witnesses tailored for different bases gives mounting evidence of its incompatibility with a quantum network.
Therefore, in the following we will extract witnesses for multiple no-input distributions, and we will evaluate each distribution in all of them to understand which distributions are easier to detect as incompatible, in the sense that they violate the largest amount of witnesses.
The main results for the state created in the network in Fig.~\ref{fig:NQKDNetwork}(b-ii) are shown in Fig.~\ref{fig:NoInputGHZ}.
There, the color code denotes the value of the witness of incompatibility of the distribution obtained by measuring the state with the operators indicating the row with the network of Fig.~\ref{fig:NQKDNetwork}(b-iii), when evaluated in the distribution obtained by measuring the state with the operators indicating the column.
More complete figures, for all possible measurement bases, can be found in Appendix \ref{app:noinput}, and all the witnesses found are stored in the computational appendix \cite{compapp}.

Since the data in Ref.~\cite{webb2023} contains statistics for all measurement choices in $\{X,Z\}^{\times 6}$, we analyze all such distributions, assessing again their compatibility with the second-order quantum inflation of the network in Fig.~\ref{fig:NQKDNetwork}(b-iii), depicted in Fig.~\ref{fig:inflation} in Appendix \ref{app:inflation}.
We obtain witnesses of incompatibility for a total of 40 distributions.
When evaluating them on the distributions resulting from considering that the sources distribute Werner states of visibility $v$, these witnesses allow to detect incompatibility for visibilities ranging between $0.7808$ (for the distribution corresponding to measurements $XZZZZX$, that establishes key between the four central parties) to $0.4094$ (for the distributions corresponding to measurements $ZXZZZZ$, $ZZXZZZ$, $ZZZXZZ$ and $ZZZZXZ$, that establish key between five of the six parties).
Then, we evaluate the witnesses on the distributions corresponding to all measurement bases in $\{X,Z\}^{\times 6}$.
In Figs.~\ref{fig:NoInputGHZTheo} and \ref{fig:NoInputGHZExp} we show, respectively, the theoretical predictions for the noiseless distributions of the form of Eq.~\eqref{eq:expstatistics} and the evaluations on the empirical data, for a subset of witnesses and distributions.
Analogous plots for all the witnesses and distributions can be found in Figs.~\ref{fig:CertificateMatrixTheoryGHZ} and \ref{fig:CertificateMatrixExpGHZ} in Appendix~\ref{app:noinput}.
These figures show that many distributions are witnessed as incompatible by a significant amount of inequalities, giving mounting evidence to the impossibility of generating them in the network of Fig.~\ref{fig:NQKDNetwork}(b-iii).

Remarkably, we observe a very good agreement between the theory and the experimental data.
This is important given that the experimental data acquisition time per  basis on a single experimental run was only \SI{60}{s} and, over the course of several iterations,  the data acquisition time totals $\sim 3.5$ hours per basis.
On average, on a single run there are on the order of $\sim20$ six-photon events.
Moreover, as in the case of Fig.~\ref{fig:twoinput}, the results are stable with regards to the amount of data used for computing the statistics.
In Fig.~\ref{fig:singleinput_all} in Appendix~\ref{app:noinput} we illustrate this stability by plotting how a selected number of inequalities detect the incompatibility of several distributions.
As we can see, even using only a fraction of the data we consistently obtain conclusive results and, when using $\sim2000$ six-photon events, the uncertainty of the violation is below the five-sigma level.
Again, we perform the equivalent analysis for the data of Ref.~\cite{pickston_nqkdtrident_2022}, corresponding to the network in Fig.~\ref{fig:NQKDNetwork}(c), in Appendix~\ref{app:trident}.

\section*{Conclusions}
In this work we report on the first device-independent analysis on quantum network structure of experimental implementations.
We generated Bell-like inequalities with quantum inflation \cite{wolfe2021qinflation,inflapyon}, a general method that can be used with any network configuration.
Using the statistics from real experiments, we ruled out the possibility that these are generated in alternative networks that contain fewer interactions despite allowing for larger physical systems.
The fact that the procedure generates Bell-like inequalities implies that the witnesses obtained can be applied to arbitrary distributions, thus not being restricted to the experiments under scrutiny in this work.

A distribution being detected by a witness is a guarantee that such distribution cannot be generated in the corresponding network, but not being detected does not imply that the distribution can be generated in the network.
Indeed, it is possible that using more constraining inflations lead to stronger witnesses.
However, this incurs in additional computational load, which quickly goes beyond standard available resources.
In the six-photon realizations that we have tested, we were able to solve each of the necessary optimization problems in less than $2.5$ minutes in a laptop.
However, going beyond an inflation level of 2 (i.e., considering two copies of each of the sources in the realization) is beyond standard computational capabilities, both memory- and time-wise.

It is possible to find violations of the inequalities using classical resources in more general networks.
For instance, many of the no-input inequalities are violated by a uniformly random bit shared among all the parties, which necessitates of a six-partite source.
Adding the ability of the parties to perform different measurements to the systems they receive may alleviate the issue.
Obtaining witnesses that do not detect classical realizations in more general networks is an important topic that is left to future work.

Other techniques for analyzing network structure currently exist in the literature, that are based on the analysis of entropies \cite{weilenmann2017} or covariances \cite{aberg2020} of the observations.
However, these typically focus exclusively in the constraints implied by the network structure, and thus produce criteria that are satisfied even by non-quantum distributions, if they are generated according to the structure dictated by the network.
In this sense, quantum inflation becomes the most suitable tool, since it allows to take into account both the network structure and the fact that the systems distributed are quantum, and the strength of both types of constraints can be tuned independently \cite[Ch. 5]{alexThesis}.

The process described is completely general, applicable to any experimental scenario and protocol.
On a more practical side, the procedure developed in this work can be used for estimating critical values for the experimental requirements of protocols, in a spirit similar to that of Ref.~\cite{abiuso2021}.
This is especially important as the traditional loopholes associated to device-independent protocols are closed, which comes at the cost of more demanding experimental requirements \cite{gu2023experimental}.
Finally, it is possible to define, with the same computational requirements, stronger characterizations with inflation that the ones we have used here by considering the so-called \textit{linearized polynomial identification} relations \cite{pozas2023}, at the expense of not being able to obtain inequalities that can detect the infeasibility of other distributions.

We applied the procedure to the analysis of concrete realizations relevant in quantum conference key agreement, showing that the data used for generating key can be recycled to also provide guarantees on the network structure.
This provides a strong motivation for developing of novel quantum information
protocols tailored to networks, which is currently a largely unexplored field \cite{lee2018towards}.
Other applications of the methods outlined in this paper could involve the verification of networks generated in programmable photonic chips, such as the ones in \cite{vigliar_error-protected_2021}.
However, in this case one must bear in mind the difficulty of closing the locality loophole.

\vspace{1em}
\noindent
\textbf{Acknowledgements}
\newline
This work is supported by the UK Engineering and Physical Sciences Research Council (Grant Nos. EP/T001011/1.), the Spanish Ministry of Science and Innovation MCIN/AEI/10.13039/501100011033 (CEX2019-000904-S and PID2020-113523GB-I00), the Spanish Ministry of Economic Affairs and Digital Transformation (project QUANTUM ENIA, as part of the Recovery, Transformation and Resilience Plan, funded by EU program NextGenerationEU), Comunidad de Madrid (QUITEMAD-CM P2018/TCS-4342), Universidad Complutense de Madrid (FEI-EU-22-06), the CSIC Quantum Technologies Platform PTI-001, the NCCR SwissMAP (grant no. 205607), and the Swiss National Science Foundation (grant number 224561).

\hypersetup{urlcolor=blue}
\bibliography{bib}

\onecolumngrid
\clearpage
\newpage

\setcounter{figure}{0}
\renewcommand{\thefigure}{S\arabic{figure}}

\appendix
\section{Experimental setup}\label{app:experimental}
The experimental realizations from which we use data are described in detail in the corresponding references, namely Ref.~\cite{webb2023} for the setup in Fig.~\ref{fig:NQKDNetwork}(b) and Ref.~\cite{pickston_nqkdtrident_2022} for the setup in Fig.~\ref{fig:NQKDNetwork}(c).
In both, a Ti:Sapph pulsed laser, with a central frequency at \SI{774.9}{nm}, feeds type-II parametric down-conversion (PDC) sources (which correspond to the leftmost boxes on Fig.~\ref{fig:NQKDNetwork}(i) in the main text).
The sources employ aperiodically poled Potassium Titanyl Phosphate (apKTP) crystals, which can produce two-photon interference visibilities up to  $98.6\pm 1.1\%$~\cite{pickston_optimised_2021}, even without filtering the photons afterwards.
Each source can be set to produce either maximally entangled photon pairs (these are denoted by the orange $\infty$ symbols), separable pairs (no symbol), or any intermediate configuration.

So as to create larger networks, individual photons from two different pairs are combined in type-II fusion gates~\cite{browne_resource-efficient_2005} (denoted by the boxes connecting two different photon sources on Fig.~\ref{fig:NQKDNetwork}).
After the required photons are fused, each photon is sent to a tomography stage (rightmost boxes in Fig.~\ref{fig:NQKDNetwork}).
Postselection on successful events, i.e. detecting a photon in one detector per tomography stage, is performed.
Both the source configuration and the number and placement of the fusion gates reflects on the causal structure of the final state, as can be seen by comparing Figs.~\ref{fig:NQKDNetwork} (i), (ii) and (iii).

\section{Inflation methods for obtaining witnesses of incompatibility}\label{app:inflation}
The witnesses of network structure that are obtained in this work are calculated using the inflation technique.
For simplicity in the notation, we present here the main ideas behind inflation methods for the case of distributions without inputs.
Adding inputs to the construction can be done in the trivial way.
For more in-depth discussions, we refer the reader to the original works  \cite{wolfe2019inflation,wolfe2021qinflation}.

\begin{figure}
    \centering
    \includegraphics[width=\linewidth]{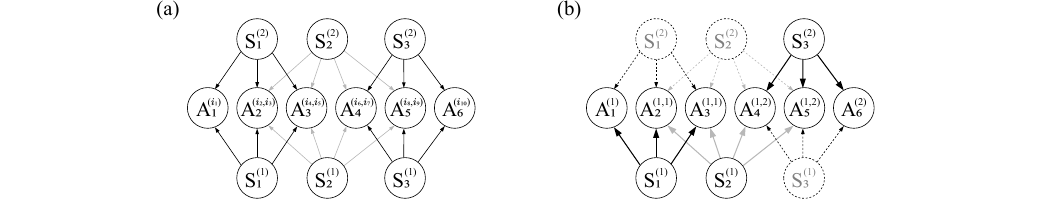}
    \caption{
        (a) Second-order quantum inflation of the network in Fig.~\ref{fig:NQKDNetwork}(b-iii).
        There are two copies of each of the sources, and each party now has access to a different copy of the original measurement operators for each combination of states they receive.
        The distributions $p_\text{inf}(\{a_1^{i_1},a_2^{i_2,i_3},\dots,a_6^{i_{10}}\}_{i_1,\dots,i_{10}})$ produced in this scenario have a number of symmetries and marginals fixed by the original distribution $p(a_1,\dots,a_6)$.
        The sequence of operators in (b) illustrate an assignment of indices ($i_1=i_2=i_4=1$, $i_3=i_5=i_6=i_8=1$, $i_7=i_9=i_{10}=2$) that reproduces the original network, and thus the corresponding marginals must reproduce $p(a_1,\dots,a_6)$.
        The fact that a $p_\text{inf}(\{a_1^{i_1},a_2^{i_2,i_3},\dots,a_6^{i_{10}}\}_{i_1,\dots,i_{10}})$ that satisfies all the necessary symmetries and marginal constraints does not exist is a proof that the premise (recall, that $p(a_1,\dots,a_6)$ can be generated in the network of Fig.~\ref{fig:NQKDNetwork}(b-iii)) is not true.
        The existence of a suitable $p_\text{inf}(\{a_1^{i_1},a_2^{i_2,i_3},\dots,a_6^{i_{10}}\}_{i_1,\dots,i_{10}})$ is a problem that can be formulated in terms of semidefinite programming \cite{npa1,npa2,tavakoli2023semidefinite}.
    }
    \label{fig:inflation}
\end{figure}

In order to demonstrate that a particular distribution cannot be generated in a given (in our case of study, quantum) network, inflation uses a strategy of reduction to the absurd: the fact that a distribution is compatible with a quantum network implies that there exist quantum states and measurement operators that reproduce it.
If such is the case, one can consider the (hypothetical) situation where access is provided to multiple copies of said states and operators, and analyze the distributions of outcomes that are produced when these are arranged in more complicated networks.
As an illustration, Fig.~\ref{fig:inflation} depicts the inflation of the network in Fig.~\ref{fig:NQKDNetwork}(b-iii) that we use throughout the main text, where we consider two copies of each of the sources of quantum systems in the original network.
There, the sources $S_i^{(k)}$ are all copies of the original source $S_i$, and the operators $A_p^{i,j}$ denote copies of the original measurement operators, $A_p$, that are applied on copies $i$ and $j$ of the corresponding sources.
If a distribution admits a realization in the network of Fig.~\ref{fig:NQKDNetwork}(b-iii), i.e., if there exist states delivered by the sources $S_1\dots S_3$ and measurement operators $A_1\dots A_6$ that reproduce said distribution via Eq.~\eqref{eq:networkstatistics}, then one can easily consider copies to produce a distribution, $p_\text{inf}(\{a_1^{i_1},a_2^{i_2,i_3},\dots,a_6^{i_{10}}\}_{i_1,\dots,i_{10}})$, in the arrangement depicted in Fig.~\ref{fig:inflation}(a).

The structure of Fig.~\ref{fig:inflation}(a) imposes a number of properties for the distributions $p_\text{inf}(\{a_1^{i_1},a_2^{i_2,i_3},\dots,a_6^{i_{10}}\}_{i_1,\dots,i_{10}})$ that can be generated in it.
First, since all $S_i^{(k)}$ represent exact copies of the same source $S_i$, expectation values of arbitrary polynomials of the measurement operators are invariant under permutation of the sources.
As an illustration (but not limited to it), one of such properties is that $p_\text{inf}(\{a_1^{i_1},a_2^{i_2,i_3},\dots,a_6^{i_{10}}\}_{i_1,\dots,i_{10}})=p_\text{inf}(\{a_1^{\pi(i_1)},a_2^{\pi(i_2),\pi'(i_3)},\dots,a_6^{\pi''(i_{10})}\}_{i_1,\dots,i_{10}})$, where $\pi$, $\pi'$ and $\pi''$ are independent permutations of the sources $S_1$, $S_2$ and $S_3$, respectively.
Second, when one restricts to marginals that reproduce the original network (or parts of it), these coincide with the original distribution under scrutiny.
This is, $p_\text{inf}(a_1^i,a_2^{i,j},a_3^{i,j},a_4^{j,k},a_5^{j,k},a_6^{k})=p(a_1,\dots,a_6)$ for any values of $i$, $j$, $k$.
Note that these are properties of any $p_\text{inf}(\{a_1^{i_1},a_2^{i_2,i_3},\dots,a_6^{i_{10}}\}_{i_1,\dots,i_{10}})$ that can be generated in the network in Fig.~\ref{fig:inflation} if the premise that $p(a_1,\dots,a_6)$ can be generated in the network in Fig.~\ref{fig:NQKDNetwork}(b-iii) is true.
Therefore, a demonstration that no such $p_\text{inf}(\{a_1^{i_1},a_2^{i_2,i_3},\dots,a_6^{i_{10}}\}_{i_1,\dots,i_{10}})$ exists is a proof that $p(a_1,\dots,a_6)$ is incompatible.

In fact, one can relax the problem to not considering distributions that can be generated in the inflated network of Fig.~\ref{fig:inflation}, but distributions that just present the required symmetries and marginals.
The set of such distributions is potentially more general, and in no case more restricted than that of the distributions that are achievable in Fig.~\ref{fig:inflation}.
Moreover, in the case of quantum distributions, this relaxed set can be characterized by a hierarchy of semidefinite programming problems, each of which is more restrictive than the next one and all of which contain the original set \cite{npa1,npa2,tavakoli2023semidefinite}.
This has two consequences that are particularly useful in our case of interest.
First, it gives a collection of tests which can be efficiently performed, and that guarantee that $p(a_1,\dots,a_6)$ cannot be generated in the network if any one of them fails.
Second, if a distribution cannot be generated in a given quantum network, and this is detected by one of the semidefinite programs in the hierarchy, it is possible to derive from this a hyperplane that separates the distribution from the set of those that pass the test.
This hyperplane is, thus, an incompatibility witness or a Bell-like inequality, that is satisfied by all compatible distributions and violated, at least, by the target distribution.
Importantly, since it is the property that provides this technique of its power in terms of certification, the violation of the inequality by any distribution is a witness of incompatibility of such distribution with the quantum network.
The opposite, namely that the inequality is satisfied, is not a guarantee of compatibility, since it could be possible that taking higher steps in the hierarchy leads to an inequality that is violated.

In order to obtain the Bell-like inequalities, we consider the distributions that are created when, in the implementation of Fig.~\ref{fig:NQKDNetwork}(b-i), the sources distribute Werner states, $\rho_v=v\ket{\phi^+}\bra{\phi^+} + (1-v)\openone/4$, where $\ket{\phi^+}=(\ket{00}+\ket{11})/\sqrt{2}$ is the maximally entangled state.
For simplicity, we assume the same visibility, $v$, for all sources.
We compute the smallest value of $v$ such that the observed correlations do not admit an inflation, and for that distribution we extract the corresponding certificate of infeasibility.
This takes the form of a polynomial Bell-like inequality, which we evaluate on the experimental data in order to detect that these, created in the setup in Fig.~\ref{fig:NQKDNetwork}(b-i), could not have been produced in the network of Fig.~\ref{fig:NQKDNetwork}(b-iii).
All this procedure is written using the \texttt{inflation} library \cite{inflapyon} and the associated codes are available in the computational appendix \cite{compapp}.

\section{Witnesses of network structure for binary-input distributions}\label{app:twoinput}
The first case we deal with is the family of distributions $p(a_1,\dots,a_6|x_1,\dots,x_6)$ that is generated when the parties measure either $\sigma_X$ or $\sigma_Z$ on their respective photons in the setup of Fig.~\ref{fig:NQKDNetwork}(b-i).
We will prove that such distribution cannot be reproduced in the quantum network of Fig.~\ref{fig:NQKDNetwork}(b-iii), without constraining neither the dimension of the states distributed by the sources nor the measurements that the parties perform on the shares they receive.
We do so by showing that $p(a_1,\dots,a_6|x_1,\dots,x_6)$ does not admit an inflation distribution in the inflation depicted in Fig.~\ref{fig:inflation}.
For this, we begin with the smallest level of the associated hierarchy of semidefinite programs.
The hierarchy is defined \cite{npa1,npa2} via sets of operators $\mathcal{O}_n$ that index the rows and columns of the matrix $\Gamma^n_{i,j}=\text{Tr}[\rho\cdot O_i^\dagger O_j]$.
If a distribution admits a quantum realization, $\Gamma^n$ is positive semidefinite for any generating set $\mathcal{O}_n$, and if it does not there exists at least one $\mathcal{O}_n$ for which $\Gamma^n$ is negative definite.
The first level of the hierarchy is defined by the set of operators $\mathcal{O}_1:=\{\openone\}\cup\{A^{i,j}_p\}$, leading to a matrix of size $41\times 41$ in our case of interest, whose positivity can be determined in $<1$ seconds.

When we assume that all the sources in Fig.~\ref{fig:NQKDNetwork}(b-i) distribute Werner states of some visibility $v$, $\rho_v=v\ket{\phi^+}\bra{\phi^+} + (1-v)\frac{\openone}{4}$ with $\ket{\phi^+}=(\ket{00}+\ket{11})/\sqrt{2}$ being the maximally entangled state, the matrix $\Gamma^1$ cannot be made positive semidefinite for $v\gtrsim 0.6180$.
This means that, at least for $v\gtrsim 0.6180$, it is not possible to reproduce the correlations generated in Fig.~\ref{fig:NQKDNetwork}(b-i) (i.e., those of the form in Eq.~\eqref{eq:expstatistics}) in the network of Fig.~\ref{fig:NQKDNetwork}(b-iii) (i.e., in the form of Eq.~\eqref{eq:networkstatistics}).
This incompatibility is witnessed by the following inequality:
\begin{equation}
    \begin{aligned}
        \mathcal{W}_1\coloneqq &\,p_{B}(0) + p_{C}(0) + p_{D}(0) + p_{E}(0) + 0.474 \left[p_{A}(0) + p_{F}(0)\right] - 0.886p_{A}(0)p_{F}(0) \\
        & -0.768 \left[p_{AB}(0,0) + p_{AC}(0,0) + p_{DF}(0,0) + p_{EF}(0,0)\right] + 0.051 \left[p_{A}(0)\left(p_{B}(0) + p_{C}(0)\right)\!+\!\left(p_{D}(0) + p_{E}(0)\right)p_{F}(0)\right] \\
        & - 0.758 \left[p_{BC}(0,0) + p_{DE}(0,0)\right] - 0.122\left[p_{B}(0)p_{C}(0) + p_{D}(0)p_{E}(0)\right] \\
        & - 0.621 \left[p_{BD}(0,0) + p_{BE}(0,0) + p_{CD}(0,0) + p_{CE}(0,0)\right] 
        + 0.041 \left[p_{C}(0) + p_{B}(0)\right]\left[p_{D}(0) + p_{E}(0)\right] \\
        & + 0.717\left[\left(p_{B}(0) + p_{C}(0)\right)p_{F}(0) + p_{A}(0)\left(p_{D}(0) + p_{E}(0)\right)\right] \\
        & - 0.031\left[p_{A}(0)^2 + p_{F}(0)^2\right] + 0.02\left[p_{B}(0)^2 + p_{C}(0)^2 + p_{D}(0)^2 + p_{E}(0)^2\right],
    \end{aligned}
    \label{eq:twoinput_1}
\end{equation}
where $p_{p_1p_2}(x_{p_1},x_{p_2})=p(a_{p_1}=0,a_{p_2}=0|x_{p_1},x_{p_2})$ is the probability of parties $p_1$ and $p_2$ obtaining outcomes $0$ when performing measurements $x_{p_1}$ and $x_{p_2}$, and the single-party probabilities are defined analogously.

Note that Eq.~\eqref{eq:twoinput_1}, despite being obtained from a two-output distribution, involves probabilities of only a single measurement, namely that denoted with the label $0$.
This indicates that the characterization provided by $\mathcal{O}_1$ is reasonably weak, and it does not exploit all the structure in the distribution.
A consequence is that one can observe violations produced by a single source distributing classical bits to the parties, since any no-input distribution can be generated in this way.
In fact, the GHZ distribution $p_\text{GHZ}(a_1,\dots,a_6)=\frac12\text{ if } a_1=\dots=a_6$ achieves a value of $\mathcal{W}_1=-0.1393$.

One can obtain a stronger witness, that exploits the full information available, by increasing the level of the hierarchy of semidefinite programs.
The next level that can be run with standard computing resources is the commonly known as level $1+\text{AB}$, defined by the subset of $\mathcal{O}_2$ given by $\mathcal{O}_{1+\text{AB}}:=\{\openone\}\cup\{A^{i,j}_p\}\cup\{A^{i,j}_p A^{i',j'}_{p'}\}_{p'\not=p}$.
Requiring that the corresponding matrix, $\Gamma^{1+\text{AB}}$ (which has size $697\times697$) is positive semidefinite (taking $\sim470$ seconds) reveals that this is impossible for $v\gtrsim 0.3887$.
The associated witness is excessively large and complex, and thus we store it in machine-readable form in the computational appendix \cite{compapp}.

\section{Witnesses of network structure for no-input distributions}\label{app:noinput}
\begin{figure}[b]
    \centering
    \includegraphics[width=\textwidth]{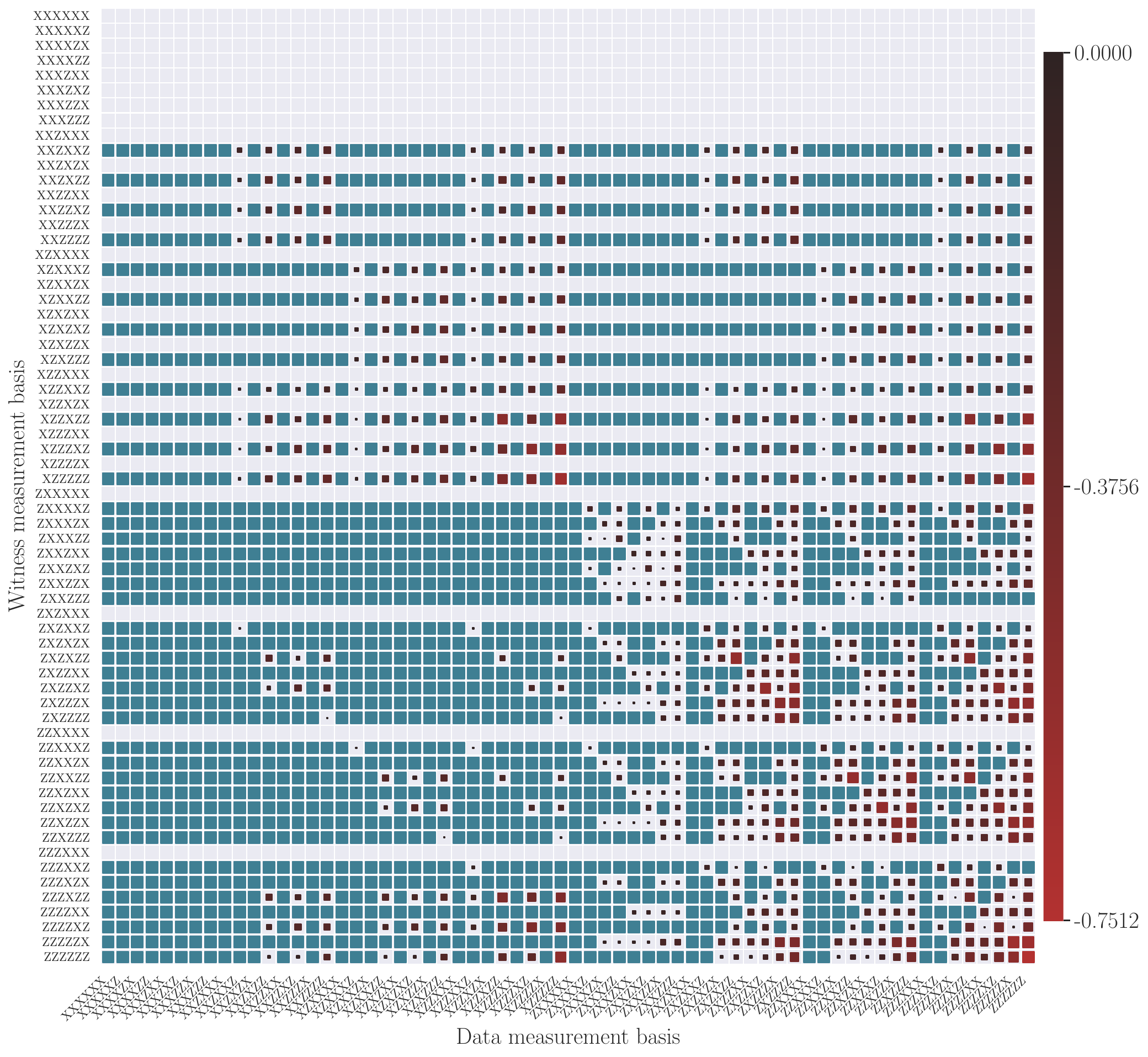}
    \caption{Analysis of the compatibility of ideal no-input distributions.
    The blue cells denote distributions that are not detected by a particular witness, i.e. those that evaluate to a positive value.
    The empty rows denote ideal distributions that are not detected to be incompatible with the inflation used.
    The size and the color of the red squares denote the strength of the detection for distributions witnessed to be incompatible.}
    \label{fig:CertificateMatrixTheoryGHZ}
\end{figure}

\begin{figure}
    \centering
    \includegraphics[width=\textwidth]{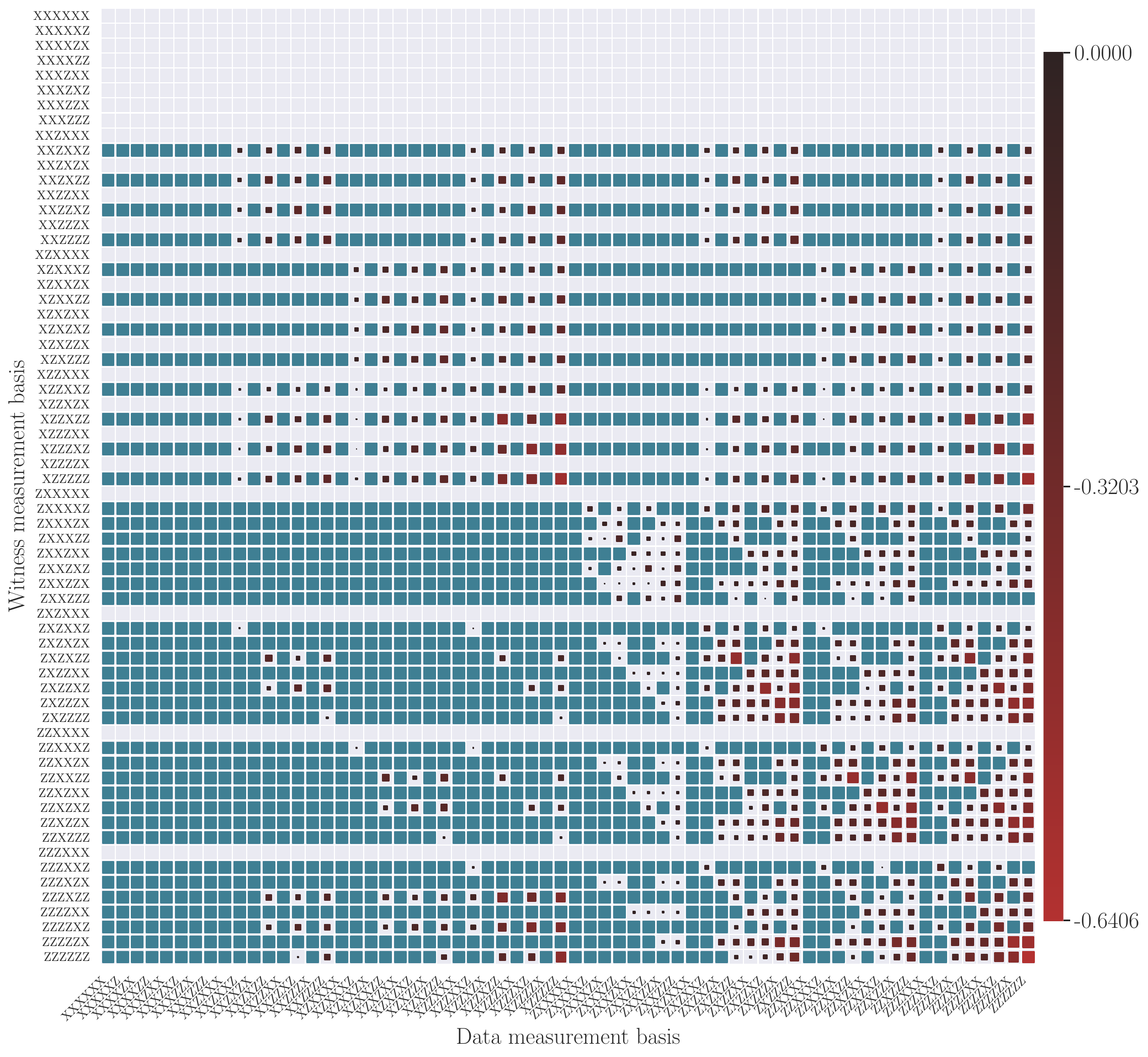}
    \caption{Analysis of the compatibility of empirical no-input distributions created from the data of Ref.~\cite{webb2023}.
    The blue cells denote distributions that are not detected by a particular witness, i.e. those that evaluate to a positive value.
    The empty rows denote ideal distributions that are not detected to be incompatible with the inflation used.
    The size and the color of the red squares denote the strength of the detection for distributions witnessed to be incompatible.
    }
    \label{fig:CertificateMatrixExpGHZ}
\end{figure}

\begin{figure}
    \centering
    \subfloat[]{
        \includegraphics[width=0.35\textwidth]{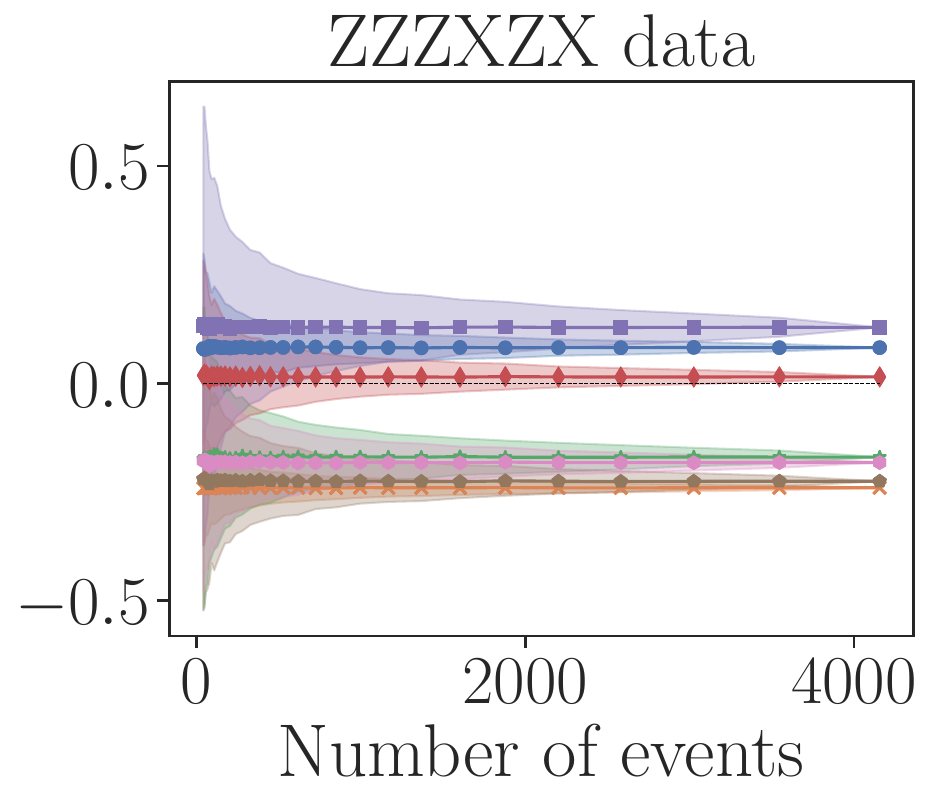}
    }
    \hskip 7em
    \subfloat[]{
        \includegraphics[width=0.35\textwidth]{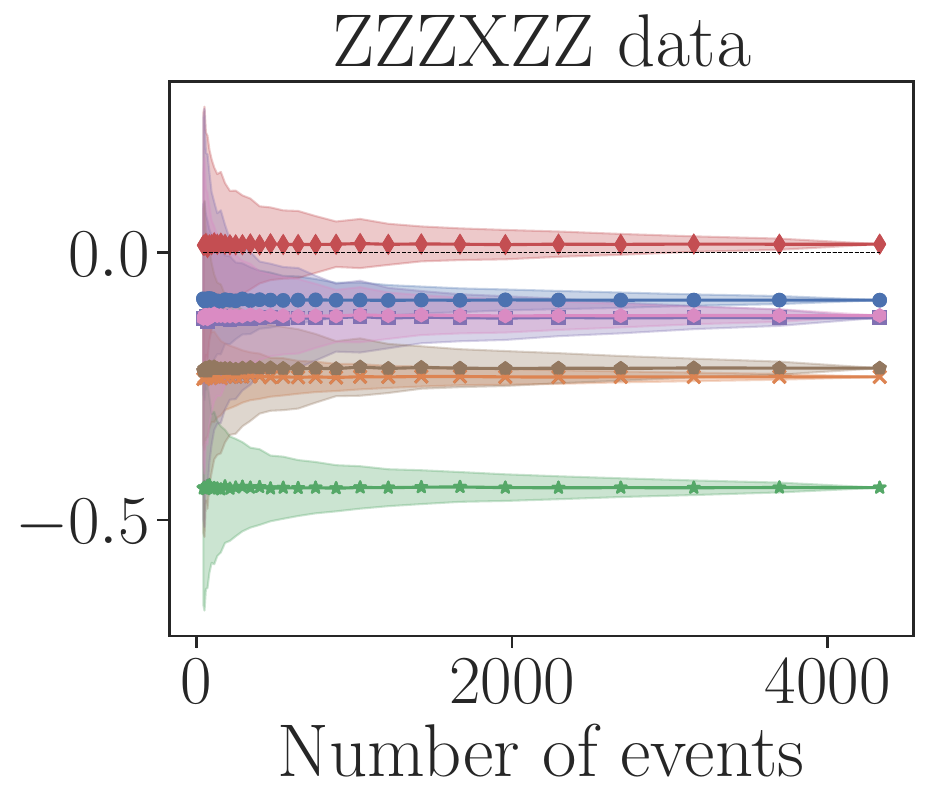}
    }
    \\
    \subfloat[]{
        \includegraphics[width=0.35\textwidth]{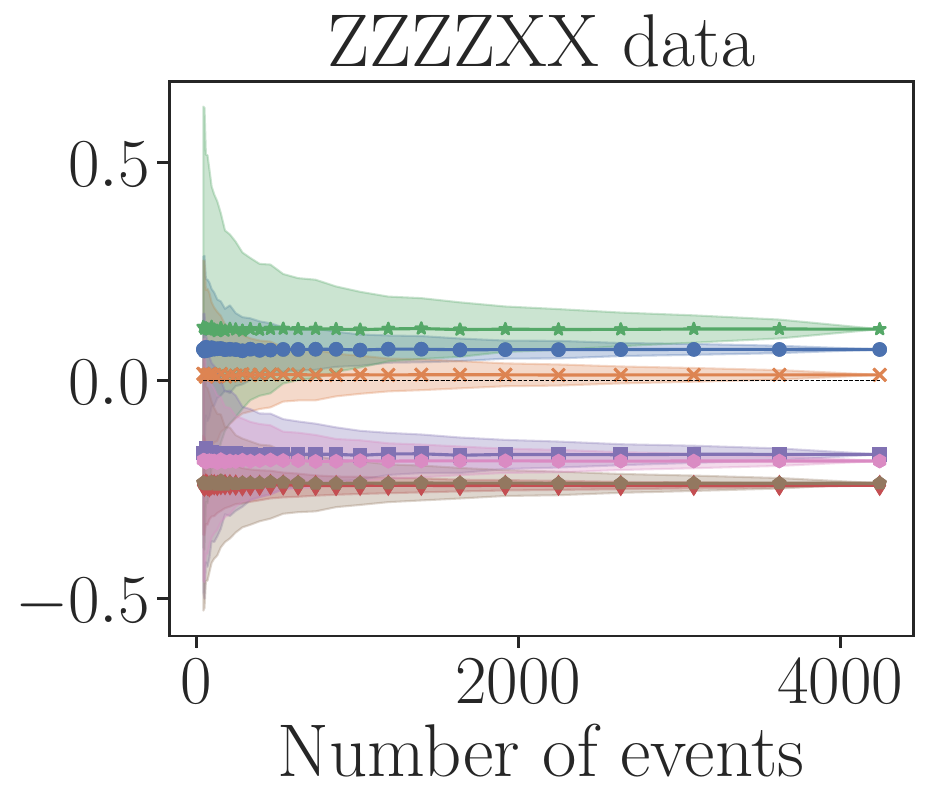}
    }
    \hskip 7em
    \subfloat[]{
        \includegraphics[width=0.35\textwidth]{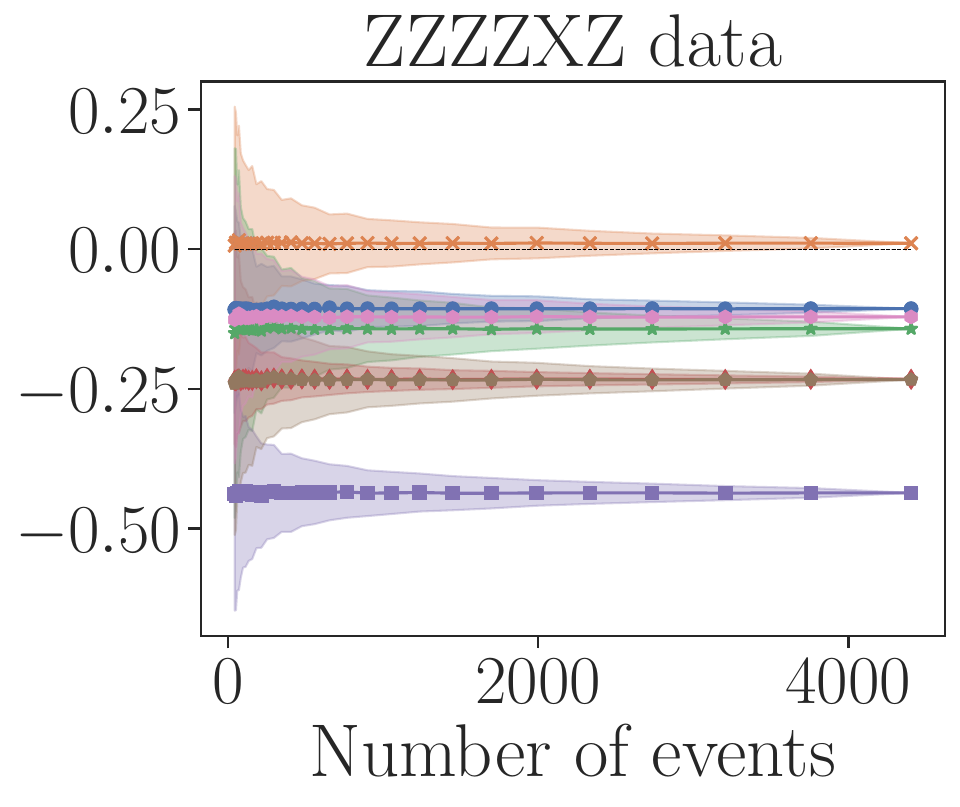}
    }
    \\
    \subfloat[]{
        \includegraphics[width=0.35\textwidth]{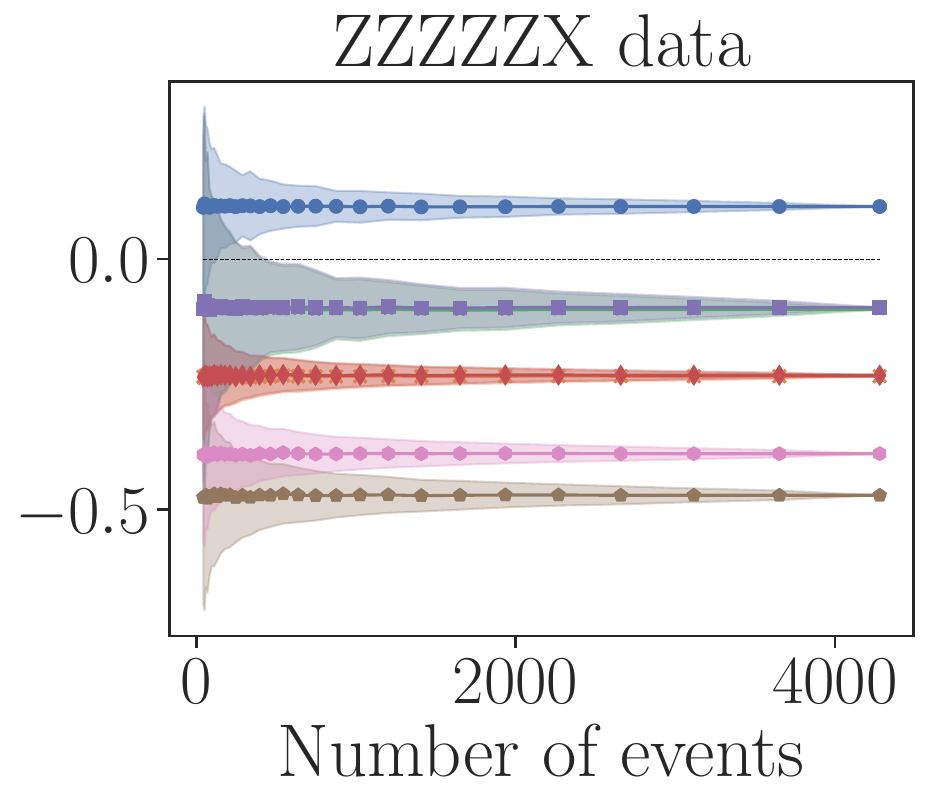}
    }
    \hskip 7em
    \subfloat[]{
        \includegraphics[width=0.35\textwidth]{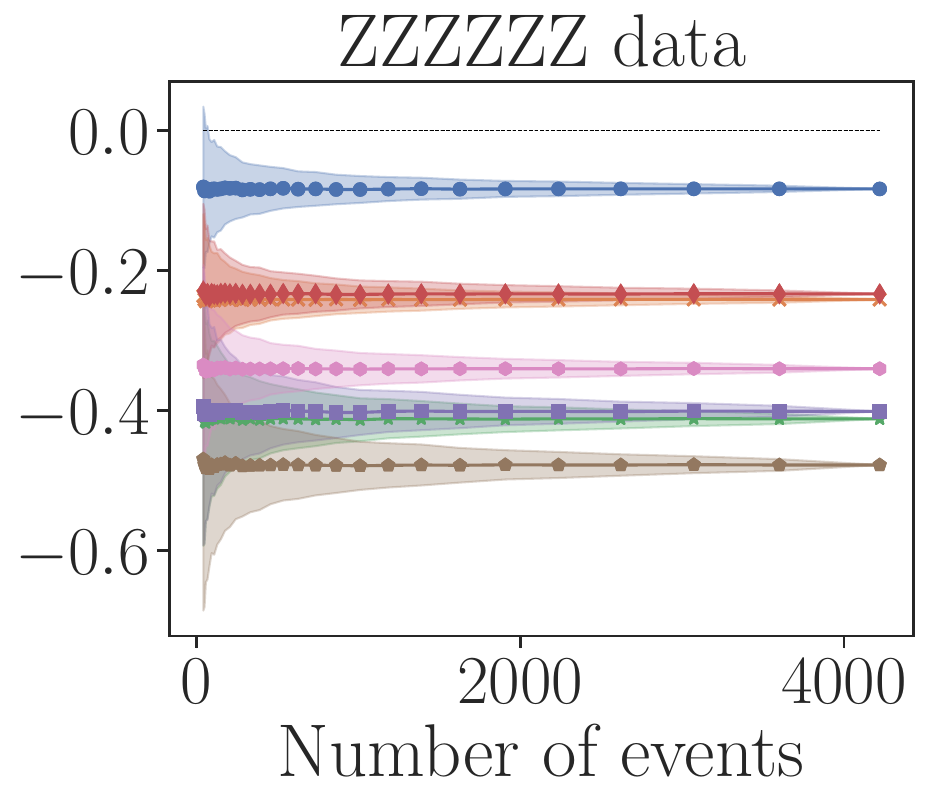}
    }
    \caption{Illustration of evaluation of the structure witnesses in the experimental data of Ref.~\cite{webb2023} for the no-input distributions as a function of the amount of experimental data used in the evaluation.
    The witnesses evaluated correspond to the following measurement bases: (blue circles) $ZZXXXZ$, (orange crosses) $ZZXXZX$, (green stars) $ZZXXZZ$, (red diamonds) $ZZXZXX$, (purple squares) $ZZXZXZ$, (brown pentagons) $ZZXZZX$, (pink hexagons) $ZZXZZZ$.
    The minimum value achievable by quantum distributions generated in the network of Fig.~\ref{fig:NQKDNetwork}(b-iii) is lower bounded by 0 in all cases (the dashed black line), so evaluating to a negative number by a given distribution is a witness that such distribution cannot be generated in Fig.~\ref{fig:NQKDNetwork}(b-iii).
    In the horizontal axis we denote the amount of all datapoints, chosen at random, used for computing the witness.
    Error bars correspond to five standard deviations over 100 repetitions.}
    \label{fig:singleinput_all}
\end{figure}

In this section we show the analysis of the compatibility of no-input distributions generated in the setup of Fig.~\ref{fig:NQKDNetwork}(b-i) with the network of Fig.~\ref{fig:NQKDNetwork}(b-iii).
The results for ideal, noiseless distributions are shown in Fig.~\ref{fig:CertificateMatrixTheoryGHZ}, and those for the experimental data of Ref.~\cite{webb2023} are shown in Fig.~\ref{fig:CertificateMatrixExpGHZ}.
In the vertical axes we write the (noiseless) distributions used to construct the corresponding witnesses, and in the horizontal axes we write the distribution that is evaluated on each of them.
The distributions are obtained by performing each of the possible sets of measurements in $\{X,Z\}^{\times 6}$ at the rightmost ports in Fig.~\ref{fig:NQKDNetwork}(b-i) when to the quantum state resulting from using maximally entangled states at the sources (leftmost ports) of the setup.
In the figures, a blank row indicates that the corresponding distribution was not detected to be incompatible with the inflation considered, so no witness could be extracted.
Figures \ref{fig:NoInputGHZTheo}, \ref{fig:NoInputGHZExp} in the main text correspond to the bottom-right corners of Figs.~\ref{fig:CertificateMatrixTheoryGHZ}, \ref{fig:CertificateMatrixExpGHZ}.
Then, Fig.~\ref{fig:singleinput_all} contains the evaluations of the witnesses that detect as incompatible the largest amount of empirical distributions, in the empirical distributions that are detected, as a function of the number of datapoints used for estimating $p(a_1,\dots,a_6)$.
All the witnesses are obtained by running the semidefinite programs corresponding to the second-order quantum inflation depicted in Fig.~\ref{fig:inflation} and the NPA level $1+\text{AB}$.
For the case of distributions without inputs, these lead to matrices $\Gamma^{1+\text{AB}}$ of size $185\times 185$, whose positivity can be determined in $\sim5.5$ seconds.

\clearpage
\section{Certification in the Trident graph}\label{app:trident}
In addition to the experimental setup of Ref.~\cite{webb2023}, we have analyzed an additional realization.
This is the one present in \cite{pickston_nqkdtrident_2022}, known as the Trident graph, and depicted in Fig.~\ref{fig:NQKDNetwork}(c).
In contrast with the previous one, the central photon source does not distribute Bell states, but separable states, and in exchange there is a third fusion gate that acts upon photons 3 and 4.
Therefore, the quantum distributions generated in it take the form

\begin{equation}
    p(a_1,\dots,a_6)=\text{Tr}\left[U_{34}\otimes U_{23}\otimes U_{45}\left(\phi^+_{12}\otimes\psi_{3}\otimes\psi_4\otimes\phi^+_{56}\right)U^\dagger_{45}\otimes U^\dagger_{23}\otimes U_{34}^\dagger\cdot\left(\Pi_{a_1}\otimes\cdots\otimes\Pi_{a_6}\right)\right],
    \label{eq:tridentstatistics}
\end{equation}
where $\psi=(\ket{0}+\ket{1})/\sqrt{2}$.

\begin{figure*}
    \centering
    \includegraphics[width=\linewidth]{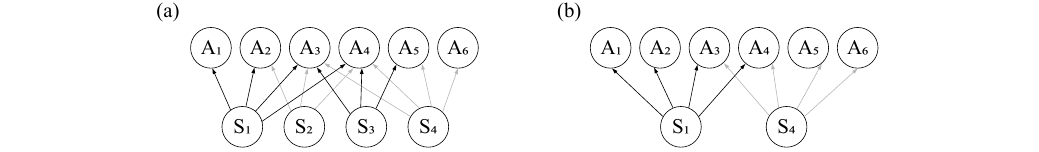}
    \caption{Closest network to the trident experiment of Ref.~\cite{pickston_nqkdtrident_2022}.
    (a) is a reproduction of Fig.~\ref{fig:NQKDNetwork}(b-iii).
    Since the dimension of the systems is unrestricted in inflation, the sources $S_2$ and $S_3$ in the naive network of (a) can be absorbed in sources $S_1$ and $S_4$, respectively, leading to the network in (b).}
    \label{fig:ExpTrident}
\end{figure*}

The associated network, obtained by connecting the sources with all the parties causally connected to them, is depicted in Fig.~\ref{fig:NQKDNetwork}(c-iii), reproduced in Fig.~\ref{fig:ExpTrident}(a).
In this case, note that the source $S_1$ distributes systems to parties $A_1$, $A_2$, $A_3$ and $A_4$, and the source $S_2$ distributes systems to parties $A_2$, $A_3$ and $A_4$.
Since in quantum inflation the dimension of the systems is not constrained, one can without loss of generality absorb the source $S_2$ into $S_1$, and the source $S_3$ into $S_4$.
Therefore, one can instead contrast against the network in Fig.~\ref{fig:ExpTrident}(b) without loss of generality.
In the following, we show results for the analysis of no-input and two-input distributions, in an analogous manner to the exposition in the main text.

\subsection{No-input distributions}
\begin{figure}
    \centering
    \subfloat[\label{fig:NoInputTridentTheo}]{
        \includegraphics[width=0.35\columnwidth]{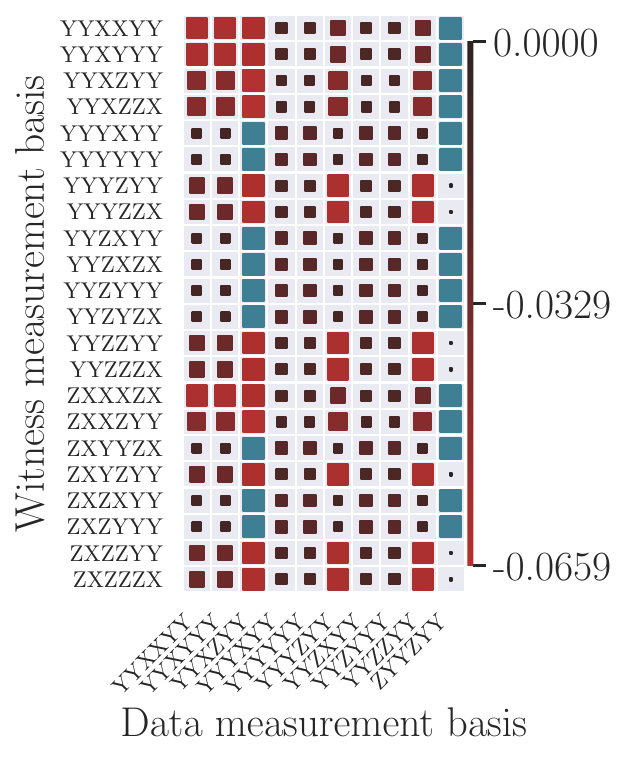}
    }
    \hspace{2em}
    \subfloat[\label{fig:NoInputTridentExp}]{
        \includegraphics[width=0.35\columnwidth]{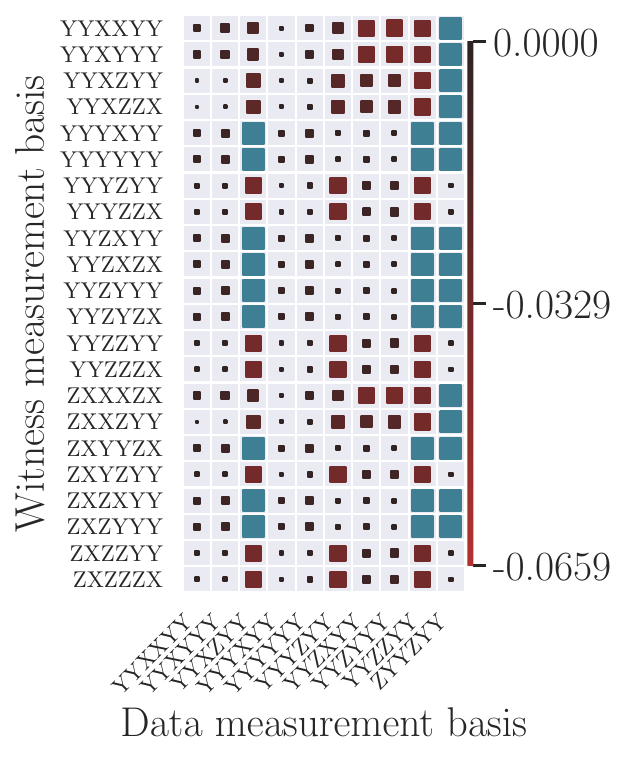}
    }
    \caption{\protect\subref{fig:NoInputTridentTheo} Theoretical predictions and \protect\subref{fig:NoInputTridentExp} experimental results for the witnesses of incompatibility with a realization of the form of Eq.~\eqref{eq:tridentnetwork}.
    The blue cells denote distributions that are not detected by a particular witness, i.e. those that evaluate to a positive value.
    The empty rows denote ideal distributions that are not detected to be incompatible with the inflation used.
    The size and the color of the red squares denote the strength of the detection for distributions witnessed to be incompatible.
    The plotted bases in the horizontal axes are the only ones for which the corresponding empirical distributions are witnessed as incompatible.}
    \label{fig:NoInputTrident}
\end{figure}

We begin by obtaining and analyzing the witnesses of no-input distributions.
Reference \cite{pickston_nqkdtrident_2022} reports data for a total of 417 choices of measurements in $\{X,Y,Z\}^{\times 6}$.
We thus assess the compatibility of all the corresponding distributions with a realization in the network of Fig.~\ref{fig:NQKDNetwork}(c-iii), i.e., of the form
\begin{equation}
    p(a_1,\dots,a_6)=\text{Tr}\left[\left(\rho_{1234}\otimes\rho_{3456}\right)\cdot\left(\Pi_{a_1}\otimes\cdots\otimes\Pi_{a_6}\right)\right].
    \label{eq:tridentnetwork}
\end{equation}
For doing so, we consider the second-order quantum inflation of Fig.~\ref{fig:NQKDNetwork}(c-iii).
The fact that this network contains one fewer source than the inflation of the network in Fig.~\ref{fig:NQKDNetwork}(b-iii) allows us to consider higher levels of the associated NPA hierarchy.
More concretely, we use the generating set $\mathcal{O}_{1+\text{AB}+\text{ABC}}:=\{\openone\}\cup\{A^{i,j}_p\}\cup\{A^{i,j}_p A^{i',j'}_{p'}\}_{p'\not=p}\cup\{A^{i,j}_p A^{i',j'}_{p'} A^{i'',j''}_{p''}\}_{p\not=p'\not=p''\not=p}$, which produces a matrix $\Gamma^{1+\text{AB}+\text{ABC}}$ of size $473\times473$, taking $\sim80$ seconds to determine its positivity.
The corresponding semidefinite programs only identify 22 distributions as not admitting a realization in terms of Eq.~\eqref{eq:tridentnetwork}.
These are the ones depicted in the vertical axes in the plots of Fig.~\ref{fig:NoInputTrident}.
When using noisy states of visibility $v$ in the sources, the incompatibility can be detected, depending on the particular witness, until visibilities ranging from $v=0.6298$ (for measurement bases $YYXXYY$, $YYXYYY$, $YYYZYY$, $YYYZZX$, $YYZZYY$, $YYZZZX$, $ZXXXZX$, $ZXYZYY$, $ZXZZYY$, and $ZXZZZX$) to $v=0.8409\sim2^{-1/4}$ (for measurement bases $YYYXYY$, $YYYYYY$, $YYZXYY$, $YYZXZX$, $YYZYYY$, $YYZYZX$, $ZXYYZX$, $ZXZXYY$, and $ZXZYYY$).

As is shown in Fig.~\ref{fig:NoInputTridentExp}, the empirical data is still in good agreement with the theory and several distributions can be witnessed as incompatible by many of the witnesses.
However, out of all the 417 empirical distributions, the only ones that are detected by the witnesses are those in the horizontal axes in Fig.~\ref{fig:NoInputTrident}.
Moreover, the relative magnitudes of the experimental to theoretical evaluations is smaller in this case than in that of the data from Ref.~\cite{webb2023} showcased in the main text.
These two phenomena may be explained by the difference in acquisition times and number of counts obtained in each experiment.
Yet, the fact that it is possible to demonstrate that the empirical distributions cannot be generated in the network of Fig.~\ref{fig:NQKDNetwork}(c-iii) motivates the research on conference key agreement protocols, such as those tested in Ref.~\cite{pickston_nqkdtrident_2022}, using the measurement bases in Fig.~\ref{fig:NoInputTrident}.

\subsection{Binary-input distributions}
For the case when the parties have a choice between two binary-outcome measurements to perform, the data available in Ref.~\cite{pickston_nqkdtrident_2022} allows to consider three simple cases, namely when all parties choose their measurements from the same pair, which can be $X-Y$, $X-Z$ or $Y-Z$.
We assess whether the corresponding theoretical distributions (i.e., in the form of Eq.~\eqref{eq:tridentstatistics}, when generalized to the parties admitting inputs) admits a realization in the form Eq.~\eqref{eq:tridentnetwork} by considering the corresponding second-order inflation, and the corresponding generating set $\mathcal{O}_{1+\text{AB}}$, that leads to assessing the positivity of a matrix $\Gamma^{1+\text{AB}}$ of size $449\times449$ (taking $\sim155$ seconds to do so).
All distributions are witnessed to be incompatible (down to visibilities of $v=0.8421$ when considering noisy states), and each of them produces one witness, namely
\begin{equation}
    \begin{aligned}
        \mathcal{W}^{1+\text{AB}}_{XY}= &\, p_{AB}(00)p_{C}(1) - p_{ABC}(001) + p_{C}(1)p_{EF}(00) - p_{CEF}(100) \\
        & + \frac{1}{\sqrt{2}}[p_{AB}(00) (p_{A}(0) + p_{B}(0) - p_{AB}(00)) + p_{EF}(00)(p_{E}(0) + p_{F}(0) - p_{EF}(00))]\\
        & + \frac{1}{2} [p_{AC}(01) - p_{A}(0)p_{C}(1) + p_{BC}(01) - p_{B}(0)p_{C}(1) + p_{CE}(10) - p_{C}(1)p_{E}(0) + p_{CF}(10) - p_{C}(1)p_{F}(0)] \\
        & + \frac{1}{2\sqrt{2}} [p_{C}(1) - p_{C}(1)^2 - p_{AB}(00) - p_{A}(0) p_{B}(0) - p_{E}(0) p_{F}(0) - p_{EF}(00)] \\
        & + \frac{1}{4\sqrt{2}}[p_{A}(0) + p_{B}(0) + p_{E}(0) + p_{F}(0) - p_{A}(0)^2 - p_{B}(0)^2 - p_{E}(0)^2 - p_{F}(0)^2],
    \end{aligned}
    \label{eq:trident_twoin_xy}
\end{equation}
\begin{equation}
    \begin{aligned}
        \mathcal{W}^{1+\text{AB}}_{XZ}= &\, p_{AB}(01) (p_{A}(0) + p_{B}(1)) + p_{EF}(01) (p_{E}(0) + p_{F}(1)) - p_{AB}(01)^2 - p_{EF}(01)^2 \\
        & + 0.471[(p_{AB}(01) + p_{EF}(01))(p_{CD}(10) + p_{C}(1) + p_{D}(0)) \\
        & \qquad\quad\,\,\, - p_{ABCD}(0110) - p_{ABC}(011) - p_{ABD}(010) - p_{CDEF}(1001) - p_{CEF}(101) - p_{DEF}(001)] \\
        & - \frac{1}{2}[p_{AB}(01) + p_{A}(0)p_{B}(1) + p_{EF}(01) + p_{E}(0)p_{F}(1)] \\
        & + 0.236[p_{ACD}(010) + p_{AC}(01) + p_{AD}(00) - p_{A}(0)(p_{C}(1) + p_{D}(0) + p_{CD}(10)) \\
        & \qquad\qquad + p_{BCD}(110) + p_{BC}(11) + p_{BD}(10) - p_{B}(1)(p_{C}(1) + p_{D}(0) + p_{CD}(10)) \\
        & \qquad\qquad + p_{CDE}(100) + p_{CE}(10) + p_{DE}(00) - (p_{CD}(10) + p_{C}(1) + p_{D}(0))p_{E}(0) \\
        & \qquad\qquad + p_{CDF}(101) + p_{CF}(11) + p_{DF}(01) - (p_{CD}(10) + p_{C}(1) + p_{D}(0))p_{F}(1)] \\
        & + \frac{1}{4}[p_{A}(0) + p_{B}(1) + p_{E}(0) + p_{F}(1) - p_{A}(0)^2 - p_{B}(1)^2 - p_{E}(0)^2 - p_{F}(1)^2] \\
        & + 0.388p_{CD}(10) \\
        & + 0.055[p_{C}(1) - p_{C}(1)^2 + p_{D}(0) - p_{D}(0)^2 - p_{CD}(10)^2] \\
        & - 0.111[p_{CD}(10)(p_{C}(1) + p_{D}(0)) + p_{C}(1)p_{D}(0)],
    \end{aligned}
    \label{eq:trident_twoin_xz}
\end{equation}
\begin{equation}
    \begin{aligned}
        \mathcal{W}^{1+\text{AB}}_{YZ}= &\, p_{AB}(00)p_{D}(1) - p_{ABD}(001) + p_{D}(1)p_{EF}(00) - p_{DEF}(100) \\
        & + \frac{1}{\sqrt{2}}[p_{AB}(00)(p_{A}(0) + p_{B}(0) - p_{AB}(00)) + p_{EF}(00)(p_{E}(0) + p_{F}(0) - p_{EF}(00))]\\
        & + \frac{1}{2} [p_{AD}(01) - p_{A}(0)p_{D}(1) + p_{BD}(01) - p_{B}(0)p_{D}(1) + p_{DE}(10) - p_{D}(1)p_{E}(0) + p_{DF}(10) - p_{D}(1)p_{F}(0)]\\
        & + \frac{1}{2\sqrt{2}} [p_{D}(1) - p_{D}(1)^2 - p_{AB}(00) - p_{A}(0)p_{B}(0) - p_{EF}(00) - p_{E}(0)p_{F}(0)]\\
        & + \frac{1}{4\sqrt{2}}[p_{A}(0) + p_{B}(0) + p_{E}(0) + p_{F}(0) - p_{A}(0)^2 - p_{B}(0)^2 - p_{E}(0)^2 - p_{F}(0)^2].
    \end{aligned}
    \label{eq:trident_twoin_yz}
\end{equation}
Note that Eqs.~\eqref{eq:trident_twoin_xy} and \eqref{eq:trident_twoin_yz} are essentially the same inequality with the role of parties $C$ and $D$ exchanged.
Moreover, as in the case with the setup of Ref.~\cite{webb2023}, the inequalities are effectively single-input inequalities.

A straightforward calculation shows that each witness is violated by the corresponding distribution, but not by those corresponding to other sets of inputs.
The evaluations on the experimental data are shown in Table \ref{tab:trident_twoinput}.
None of the evaluations give a clear violation, and thus one needs to consider higher levels in the semidefinite programming hierarchy in order to obtain witnesses that are violated by the empirical data.

\begin{table}[h]
    \begin{tabular}{c|r|r|r|r|r|r}
        & \multicolumn{1}{c}{$XY$} & \multicolumn{1}{c}{$XZ$} & \multicolumn{1}{c}{$YX$} & \multicolumn{1}{c}{$YZ$} & \multicolumn{1}{c}{$ZX$} & \multicolumn{1}{c}{$ZY$} \\
        \hline
        $XY$ & $0.1687\pm0.0240$ & $0.3720\pm0.0278$ & $0.1759\pm0.0251$ & $-0.0137\pm0.0294$ & $0.1164\pm0.0320$ & $0.1125\pm0.0279$ \\
        $XZ$ & $0.2045\pm0.0231$ & $0.1929\pm0.0266$ & $0.2000\pm0.0227$ & $0.1978\pm0.0281$ & $0.1910\pm0.0284$ & $0.2065\pm0.0269$ \\
        $YZ$ & $0.1809\pm0.0271$ & $0.3695\pm0.0275$ & $0.1779\pm0.0268$ & $-0.0050\pm0.0317$ & $0.1105\pm0.0293$ & $0.1103\pm0.0283$
    \end{tabular}
    \caption{Evaluations of the witnesses in Eqs.~\eqref{eq:trident_twoin_xy}-\eqref{eq:trident_twoin_yz} (in the rows) in the experimental data of \cite{pickston_nqkdtrident_2022} (in the columns).
        The order of the bases in the first row determines which basis corresponds to measurements $0$ and $1$ in Eqs.~\eqref{eq:trident_twoin_xy}-\eqref{eq:trident_twoin_yz}.}
    \label{tab:trident_twoinput}
\end{table}

\end{document}